\documentstyle[epsfig,epsf,12pt,a4]{article}

\newcommand{\beq}{\begin{equation}}
\newcommand{\eeq}{\end{equation}}
\newcommand{\bdm}{\begin{displaymath}}
\newcommand{\edm}{\end{displaymath}}

\begin{document}


\title{\bf{Coherent Photoproduction of Dileptons on Light Nuclei - 
a New Means to Learn about Vector Mesons}}

\author{M. Post , W. Peters and U. Mosel \\
        Institut f\"ur Theoretische Physik, Universit\"at Giessen \\
        D-35392 Giessen, Germany}

\date{}

\maketitle


\begin{abstract}

In the last years much work has been done to learn about the properties of
the $\rho$-meson in nuclear matter. In a calculation of the in-medium
spectralfunction of the $\rho$ we find that it is strongly
modified by baryonic resonances, especially the $D_{13}(1520)$, through its coupling
to resonance-hole states. In order to test the predictions of a model for
the in-medium properties of the $\rho$-meson so far mostly heavy-ion collision have
been used. We argue that the coherent photoproduction of $\rho$-mesons 
yields additional information about the nature of the medium modification. 
It can be shown
that the production amplitude is sensitive on the momentum dependence of
the $\rho$-selfenergy. This can be used to distinguish experimentally
between various models for the $\rho$ in matter.   

\end{abstract}


\section{Introduction}
\label{intro}

The question of how the $\rho$-meson behaves in hot and dense nuclear matter has 
attracted much attention over the last years. Based on arguments from chiral
symmetry, one expects that the $\rho$-meson mass changes in the vicinity of 
the chiral phase transition, though chiral symmetry does not tell if its mass
goes up or down \cite{ko95}. 
The interest in this question was further stimulated 
by measurements of dilepton spectra from heavy-ion collions,
which were carried out by the NA45 collaboration \cite{ceres2}. 
These spectra seem to indicate
a mass-shift of the $\rho$-meson down to lower masses of about $100$ MeV.

In order to understand this effect various theory groups 
\cite{brownrho,herrmann,rapp,kw97,friman,frlu98,pepo} performed calculations
of the selfenergy of the $\rho$-meson at finite density, which
contains all the information about its mass and decay width in nuclear matter. 
The models differ a lot in what they predict for the in-medium properties 
of the $\rho$-meson, the results ranging from a mass-shift of the 
$\rho$ \cite{brownrho}to a selfenergy, which clearly shows resonant 
structures from the 
excitation of baryonic resonances, especially the 
$D_{13} (1520)$ \cite{pepo}.
However, if one uses these selfenergies to calculate dilepton spectra 
in heavy-ion collisions
it turns out that they yield very similar results \cite{cb97}
and thus it seems to be very 
hard to distinguish experimentally between them on the basis of heavy-ion collions.

Therefore it is clearly necessary to find other reactions that yield additional
information about the $\rho$-meson in medium. We claim that the photoproduction
of $\rho$-mesons is a promising candidate for that. 


In this talk we will concentrate on a discussion of the coherent photoproduction
of $\rho$-mesons off light nuclei. The term coherent will be explained in section
\ref{coherent}. As we will show, this reaction is not only very sensitive to
different medium-modifications of the $\rho$-meson, in addition it also opens
up the possibility to study the momentum dependence of the selfenergy of the
$\rho$.

The talk is organized as follows: in section \ref{rhospec} we will briefly 
review the influence of the excitation of resonance-hole loops on the
$\rho$-selfenergy. 
In section \ref{coherent} we will explain the model to calculate
the coherent photoproduction before we then turn to the results.


\section{The Selfenergy of the $\rho$-meson in Nuclear Matter}
\label{rhospec}

For the study of the mass spectrum of a particle it is convenient to 
introduce the spectralfunction, which is proportional to the imaginary 
part of the propagator of the particle. Thus, for a $\rho$-meson in 
vacuum, the spectralfunction has the following form:
\bdm
\label{spec_free_def}
     A_\rho^{vac}(q) = \frac{1}{\pi} \frac{\rm{Im}\Sigma^{vac}(q)}
                    {(q^2-m_\rho^2)^2 + (\rm{Im}\Sigma^{vac}(q))^2} \,,
\edm
where
\bdm
\label{sigma_free_def}
    \rm{Im}\Sigma^{vac}(q) = \sqrt{q^2}\,\Gamma_{\rho\pi\pi}
\edm
describes the decay of a $\rho$-meson into two pions.
The spectralfunction gives the probability that the $\rho$-meson propagates
with a
mass $m = \sqrt{q^2}$. One sees, that through the coupling to the 
2$\pi$-channel the $\rho$-meson can propagate with any mass larger than 
$2\,m_\pi$ and not only 
with its rest mass of $m_\rho = 0.768$ MeV.

In nuclear matter there will be additional contributions to the selfenergy
from interactions of the $\rho$ with the surrounding nucleons:
\bdm 
    \Sigma(\omega,\vec q\,) = \Sigma^{vac}(q) + \Sigma^{med}(\omega,\vec q\,)\,.
\edm
Note that the in-medium part of the selfenergy depends on energy and three-momentum
of the $\rho$-meson independently. 
This is a direct consequence of the fact, that there exists
a preferred rest frame, namely the rest frame of nuclear matter. 
Energy and momentum 
of the $\rho$-meson are defined with respect to that frame.
As a further consequence, transversely and longitudinally polarized $\rho$-mesons
will be modified differently.  

At low nuclear densities the $\rho$-selfenergy can be calculated by means 
of the {\it low density theorem}, which relates the selfenergy to the
$\rho\,N$ forward-scattering amplitude:
\bdm
     \Sigma^{med} = \rho \, {\cal T}_{\rho\,N} ({\theta} = 0)
\edm

Thus, at low densities a complete knowledge of the 
$\rho\,N$ forward-scattering amplitude suffices for a description
of the $\rho$-mass spectrum in nuclear matter.
There are various contributions to this amplitude. In this talk we want to 
concentrate on those scattering processes that lead to the excitation of
a baryonic resonance (fig.\ref{bar_scatt_picf}):

\begin{figure}[h]
\begin{center}
\centerline{\epsfxsize=8cm \epsfbox{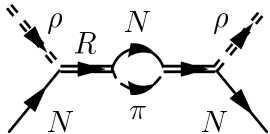}}
\caption{\label{bar_scatt_picf} Contribution to the $\rho\,N$ scattering
amplitude that leads to the excitation of a baryonic resonance.}
\end{center}
\end{figure}

The details of the calculation can be found in Peters {\it et al.} 
\cite{pepo}, to which 
the interested reader may refer. In addition, we would like to discuss here
some points that have not been mentioned in our previous publication.

If one looks up baryonic resonances which couple to the $\rho\,N$-channel
in \cite{pdg}, one finds some resonances  
whose mass $m_R$ is below the $\rho\,N$-threshold:
\bdm
     m_R < m_\rho + m_N \, .
\edm

Among these resonances is for example the $D_{13}(1520)$-resonance which
in \cite{pepo} was found to be very important for the in-medium properties
of the $\rho$-meson. However, keeping in mind that the $\rho$ is an 
unstable particle, this is not puzzling at all: the resonances simply 
couple to the low-mass tail of the $\rho$-spectralfunction.

\begin{figure}[h]
\centerline{\epsfxsize=8cm \epsfbox{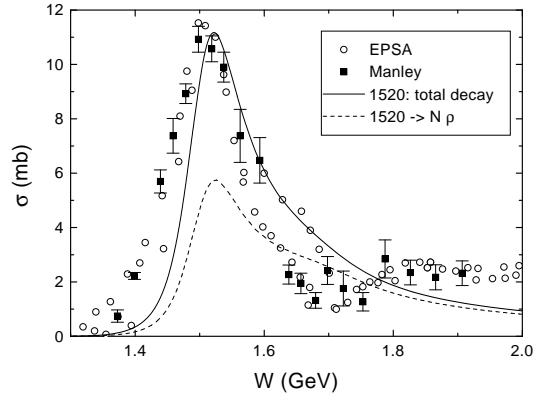}}
\caption{\label{manley_d13_fig} Results for the $D_{13}$-channel from
a partial-wave analysis of $\pi\,N \rightarrow \pi\,\pi\,N$-data by Manley
\cite{manley}.
}
\end{figure}

Direct experimental evidence for that can be found in an analysis from
Manley {\it et al.} \cite{manley}. He performed a partial-wave analysis of all
existing data for the reaction $\pi\,N \rightarrow \pi\,\pi\,N$ within
an isobar model, allowing for $\rho\,N$, $\Delta\,\pi$ and $\epsilon\,N$
as intermediate $\pi\,\pi\,N$-states. Here the $\epsilon$ represents
an isoscalar $s$-wave $\pi\,\pi$-state.
Because of its importance we want
to discuss the case of the $D_{13}(1520)$-resonance. 
The result of the analysis for the corresponding partial-wave together with 
the contribution from the coupling of the resonance to $\rho\,N$ is shown in 
fig.\ref{manley_d13_fig} and leaves little doubt that the $D_{13}(1520)$
really decays into $\rho\,N$.

\begin{figure}[h]
\centerline{\epsfxsize=8cm \epsfbox{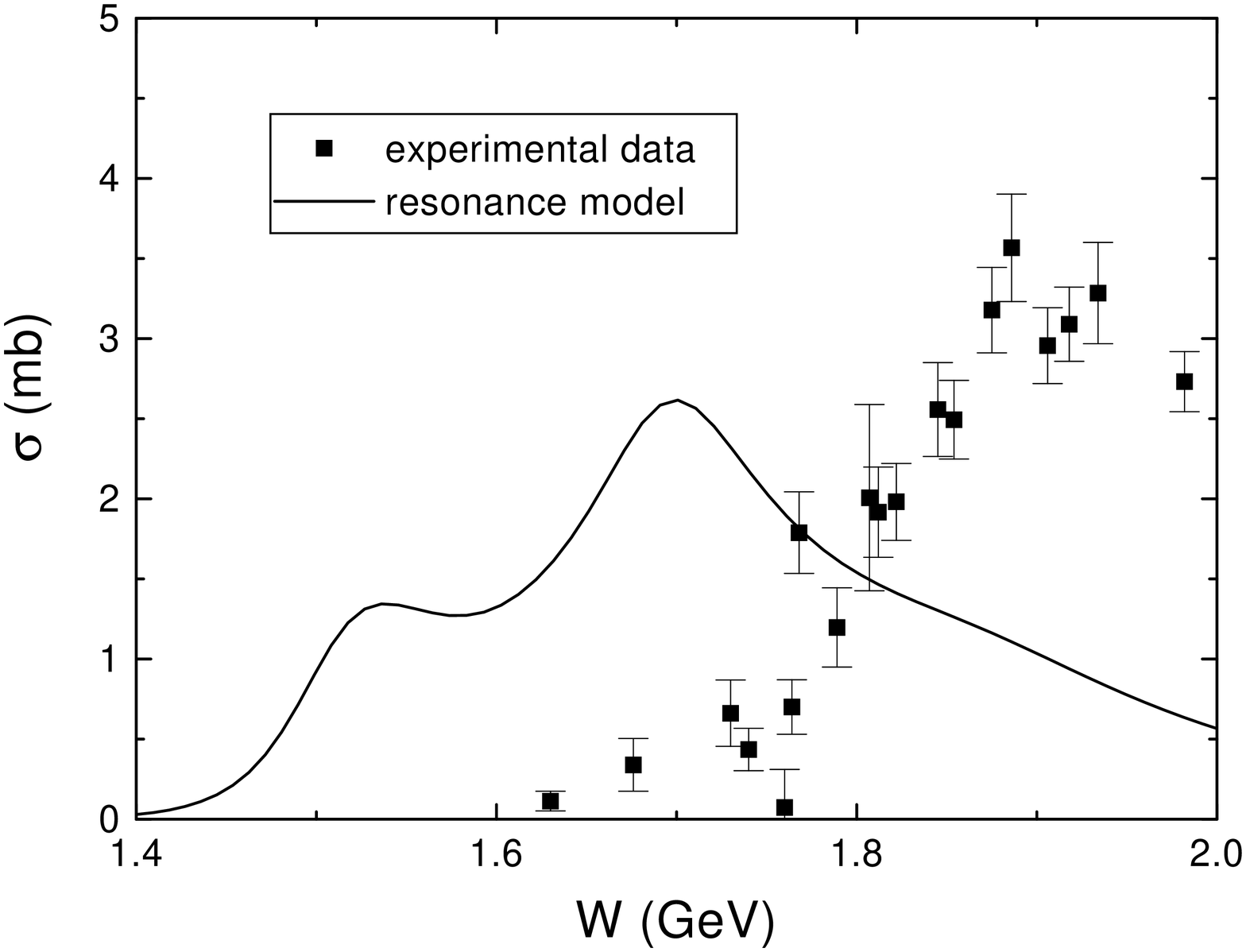}}
\caption{\label{brody_fig} Experimental data for the reaction 
$\pi^-\,p \rightarrow \rho^0\,n$ shown in comparison with 
a calculation based on the resonance-hole selfenergy. The data are
taken from \cite{brody}}
\end{figure}

We mentioned before that the knowledge of the $\rho\,N$ forward-scattering
amplitude suffices at low densities for a complete description of the
in-medium properties of the $\rho$-meson. As was pointed out by Friman 
\cite{fr98}, parts of this amplitude can be compared with 
the experimental data for the reaction $\pi^-\,p \rightarrow \rho^0\,n$.
The only available analysis of this is reaction is from Brody 
{\it et al.} \cite{brody}.
It is shown in fig.\ref{brody_fig} in comparison with a calculation of the
cross-section based on our $\rho\,N$-scattering amplitude.

The calculation seems to be in clear contradiction to the data. We argue however,
that this does not imply that the used model is incorrect, but rather that
the extraction of the data at energies $W < 1.7$ GeV is not 
reliable. The data suggest, for example, that there is no coupling of 
the $D_{13}(1520)$ to $\rho\,N$ which, as shown above, does not agree
with Manleys analysis. The latter was carried out more carefully 
and is based on a much larger set of data, so we prefer to rely on its results. 
At higher energies the agreement
becomes better, due to the fact that the experimental identification of the 
$\rho$-mesons is less problematic.

Before we turn to the coherent photoproduction let us quickly review
one main feature of the spectralfunction resulting from the selfenergy
discussed above. The calculations show that transverse and longitudinal 
selfenergy have a very different momentum dependence.
At low momenta both exhibit clearly the influence of the $D_{13}(1520)$.
However, whereas the spectralfunction for transverse $\rho$-mesons 
is nearly flat at high momenta, for longitudinal $\rho$-mesons
the importance of the $D_{13}(1520)$ as well as of the other resonances 
is reduced and much of the
structure coming from the $\rho$-decay into pions can be found. This 
will be of great importance in the next section.


\section{Coherent Photoproduction of Vector-Mesons}
\label{coherent}

We come now to the discussion of the coherent photoproduction of 
$\rho$ mesons off light nuclei. By coherent we mean that the 
$\rho$ is produced elastically, i.e. the nucleus is required to remain
in its ground state. Since the $\rho$-meson has a large decay width,
it will decay inside the nucleus. In order to avoid a distortion of the
signal due to final state interactions of the decay products, we consider
dileptons as the final state. One also has to calculate the dilepton
production rate coming from intermediate photon and
$\omega$-meson states, wich can not be distinguished experimentally from
dileptons from coming $\rho$-decay.

In impulse approximation the amplitude for the complete process can be put in 
the form:

\bdm
    {\cal M} \sim \sum_V \int \! dm \sum_{\alpha} \frac
                 {\langle e^+e^- | {\cal O} | V(m) \rangle
                 \langle \alpha \, V(m)| {\cal V}
                        |\alpha \, \gamma \rangle}
                 {m^2-m_V^2 + I\,m\,\Gamma + \Sigma^{med}} 
\edm

Here $V$ represents the produced spin-1 state, $m_V$ its mass,
$\Gamma$ its vacuum decay width and
$\Sigma^{med}$ its selfenergy in nuclear matter.
Different scenarios for the medium-modifications of the $\rho$ enter through
$\Sigma^{med}$. 
$m$ is the invariant mass of the dileptons. $ |\alpha  \rangle$
is a bound nucleon state with the quantum numbers $\alpha$ and the sum
is over all filled nucleon states in the nucleus under consideration.
The potential
for the production of a vector-meson is denoted by $ {\cal V}$ and ${\cal O}$
describes the coupling of a vector particle to dilepton.


\subsection{The Potential ${\cal V}$}

The potential ${\cal V}$ is taken from Friman {\it et al.} \cite{fs97}
and describes the photoproduction of vector-mesons within a meson-exchange model.
The parameters of the model are adjusted to data for the photoproduction
on free nucleons. It turns out that for a
reasonable description of the $\rho$-meson production 
one needs to take into account the contribution from both
$\pi$- and $\sigma$-exchange, whereas in the case of $\omega$-mesons 
$\pi$-exchange alone suffices. Since the pion is a pseudoscalar particle,
$\pi$-exchange
induces a change of the parity of the nucleus and does therefore
not contribute to the amplitude for the coherent production.
Thus within our model this amplitude vanishes for $\omega$-mesons.


\subsection{The Selfenergy $\Sigma^{med}$}

The selfenergy $\Sigma^{med}$ describes how the $\rho$-meson is modified during its
propagation through the nucleus. In our calculation we studied
the effects of a selfenergy based on the excitation of resonance-hole loops, which
was discussed in the first part of this talk, on the production amplitude.
We would like to mention again the main properties of this model, namely
that it has a large imaginary part and that it shows a different 
momentum-dependence of transverse and longitudinal selfenergy.

\begin{figure}[h]
\begin{center}
\centerline{\epsfxsize=7cm \epsfbox{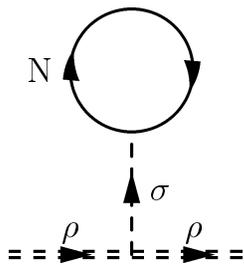}}
\caption{\label{tadpole_fig} Tadpole contribution to the $\rho$-selfenergy
that follows from the potential ${\cal V}$.}
\end{center}
\end{figure}

In order to demonstrate the sensitivity of the amplitude to different models
for the in-medium modification of the $\rho$-meson we also calculated
the $\rho$-selfenergy that follows from the same Lagrangian as the 
potential ${\cal V}$ \cite{fs97} and which is depicted diagramatically 
by a tadpole-graph (fig.\ref{tadpole_fig}).

In contrast to the resonance-hole model this selfenergy is purely real
and leads to a decrease of the $\rho$-mass of about $100$ MeV. Also,
it induces the same medium-modification to both transverse and longitudinal
$\rho$-mesons.


\subsection{Results}

\begin{figure}[h]
\centerline{\epsfxsize=9cm \epsfbox{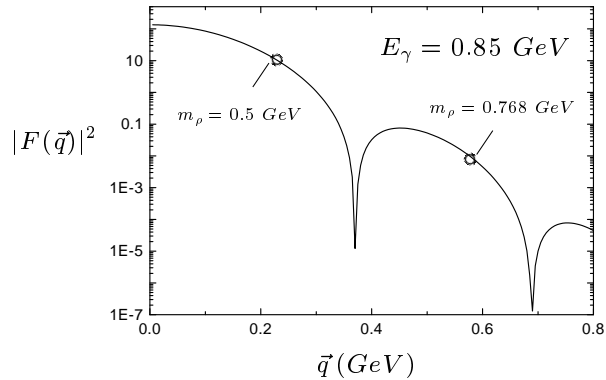}}
\caption{\label{formfac_fig} The nuclear formfactor of $^{12}C$. Also indicated
are the values of $q_{min}$ for the production
of a particle of mass $0.5$ GeV and $0.768$ GeV.}
\end{figure}

The calculation shows that with our choice of the potential ${\cal V}$
the production amplitude is proportional to the nuclear formfactor $F(q)$ ,
where $q$ denotes the momentum transfer. In fig.\ref{formfac_fig} we show
the formfactor of $^{12}C$. We also indicate the minimal momentum transfer
$q_{min}$ for the production of a particle of mass $0.5$ GeV and $0.768$ GeV
at an incident photon energy of $0.85$ GeV.

A simple kinematical consideration shows that with dropping mass 
or increasing photon energy $q_{min}$ becomes smaller.

In general
\bdm
 \sigma_{tot} \,,\,\frac{d\sigma}{dm} \propto 
 \int_{q_{min}}^{q_{max}}\,q |F(q)|^2 \, .
\edm
Since the formfactor decreases rapidly as $q$ increases, it is clear
that the magnitude of the cross-section is mainly determined by the 
kinematical region around $q_{min}$. As a direct consequence of the 
kinematics the nuclear formfactor will therefore strongly favour
the production of $\rho$-mesons lighter than $m_\rho = 0.768$ GeV, 
whereas the spectralfunction favours $\rho$-mesons with a mass around
$m_\rho$. Thus one expects that the shape of $\frac{d\sigma}{dm}$
is governed by an interplay between spectralfunction and formfactor
and that two peaks will show up in the spectrum. 

For the same reason 
the coherent photoproduction is very sensitive to medium modifications of the
$\rho$-meson. The cross-section for a $\rho$-meson whose mass is reduced
in the nuclear medium will be substantially larger than in the vacuum-case.
If on the other side the major effect of the medium is a broadening 
of the $\rho$, the cross-section should be reduced due to absorptive effects.  

\begin{figure}[top]
\centerline{\epsfxsize=12cm \epsfbox{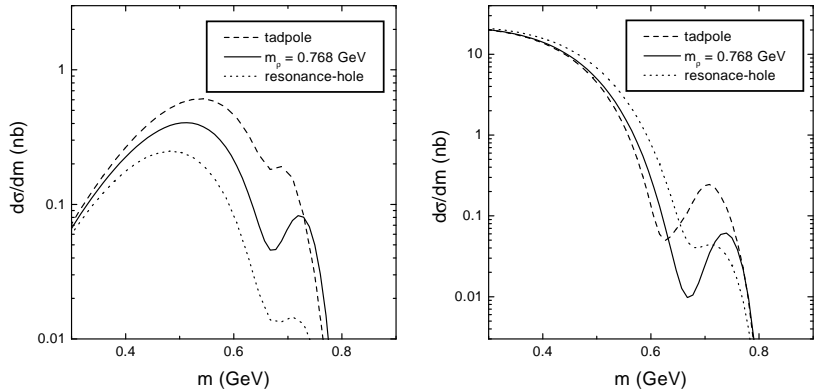}}
\caption{\label{dm_both_fig} $\frac{d\sigma}{dm}$ of the reaction
 $\gamma \,^{12}C \rightarrow V \, ^{12}C \rightarrow e^+e^-\,^{12}C$ 
for various medium-modifications of the $\rho$-meson. The plot on the left
shows the contribution from the $\rho$ alone while on the right one sees the
contribution from both $\rho$ and photon.}
\end{figure}

In fig.\ref{dm_both_fig} $\frac{d\sigma}{dm}$ for the production of dileptons
via vector-mesons is shown for different
medium-scenarios at a photon energy of $0.85$ GeV. The left plot contains
only the contribution from the $\rho$-meson to the dilepton-spectrum.
The results are in line with the previous discussion. Two
peaks can be found in the spectrum at $m \sim 0.77$ GeV and at $m \sim 0.55$ GeV.
Furthermore, a lower $\rho$-mass stronlgy enhances the cross-section
whereas the resonance-hole model, which predicts a strong broadening
of the $\rho$, gives smaller results. The plot on the right shows 
$\frac{d\sigma}{dm}$ with the photon included as well.
The spectralfunction of the photon enhances the contribution at low masses.
Besides, the decay width of a virtual photon
into dileptons has a different mass dependence than that of a
$\rho$-meson ($\Gamma_\gamma \sim \frac{1}{m^3}$, but $\Gamma_\rho 
\sim m$). Both effects lead to a strong enhancement of the cross-section
at low masses. However, various medium-modifications of the $\rho$-meson
still lead to quite different results for masses above $0.6$ GeV.

\begin{figure}[h]
\centerline{\epsfxsize=9cm \epsfbox{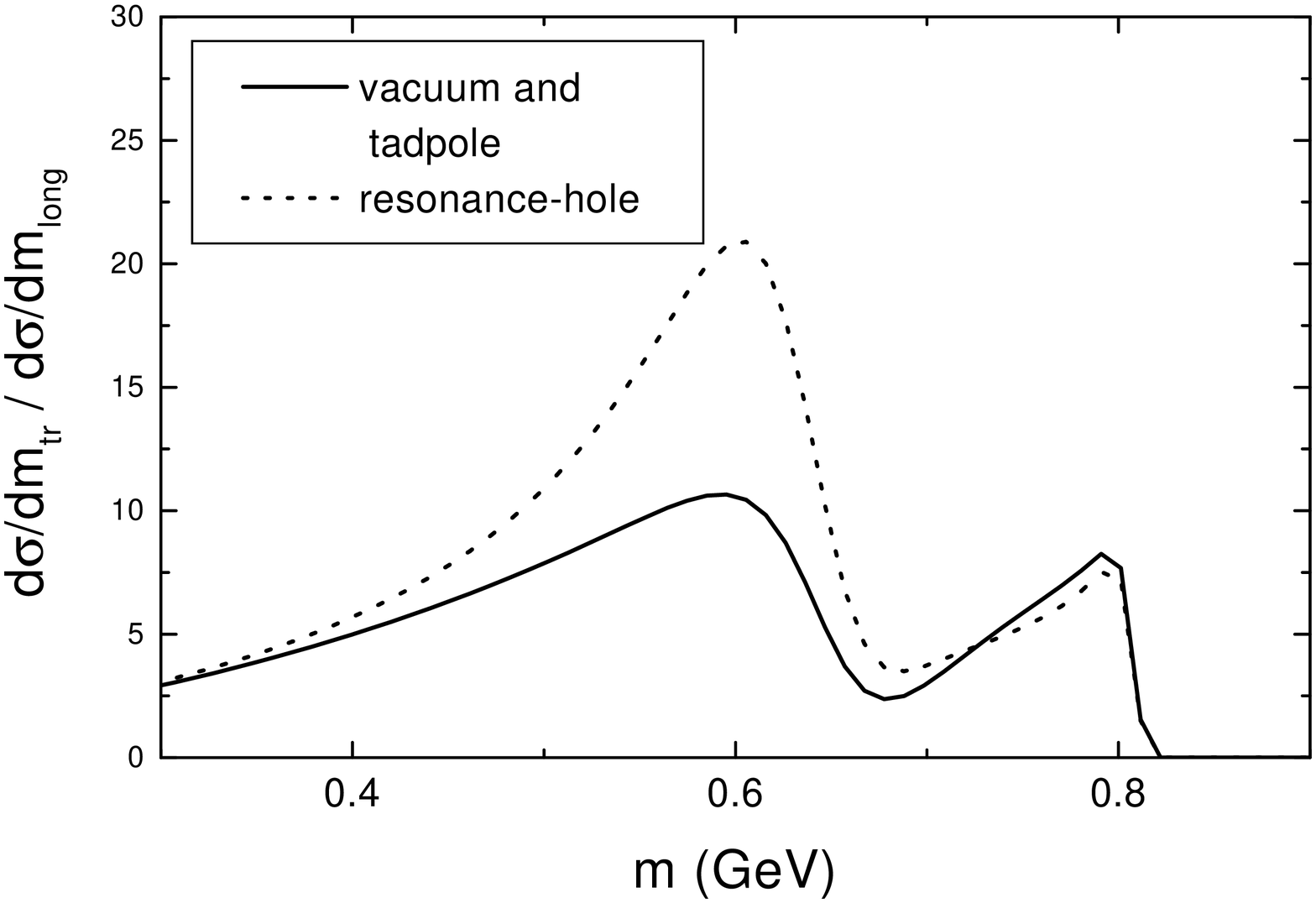}}
\caption{\label{dm_ratio_fig} The ratio $R$ as defined in the text plotted
for different medium-modifications of the $\rho$-meson.}
\end{figure}

The results shown so far did not take into account the polarization 
of the vector-particles. In view of the very different momentum dependence
of the selfenergy for transverse and longitudinal $\rho$-mesons
in the resonance-hole model it is tempting to look at both polarizations
separately and thus to turn the momentum dependence directly into an
observable. We find that the ratio 
$R = \frac{d\sigma_{tr}}{dm}/\frac{d\sigma_{long}}{dm}$ is of particular interest.
In fig.\ref{dm_ratio_fig} we show $R$ for the two selfenergies discussed above
and for the vacuum case. The resonance-hole model leads to a strong enhancement
of $R$ in the mass region around $0.6$ GeV in comparison to the vacuum case.
Since the tadpole-selfenergy is identical for transverse and longitudinal 
polarizations and since $R$ is proportional to the ratio of both selfenergies,
a simple mass-shift scenario gives exactly the same results as one would get
for a $\rho$-meson without any medium modification. 

\section*{Summary $\&$ Outlook}

In a calculation of the $\rho$-selfenergy in nuclear matter 
we found that the excitation
of the $D_{13}(1520)$-resonance in $\rho\,N$ scattering is of great importance
for the in-medium properties of the $\rho$-meson. As a possibility to 
obtain more information about the $\rho$ in nuclear matter we propose the
coherent photoproduction of vector mesons. It was demonstrated that
the production rates are quite sensitive to different in-medium
scenarios for the $\rho$-meson. Furthermore, by looking at the polarization
of the vector-meson one can obtain valuable information about the
momentum dependence of the selfenergy of the $\rho$. 

Further work 
on this subject will include the calculation of a background contribution,
the Bethe-Heitler process \cite{tsai}, 
to the dilepton spectrum, a more refined
version of the potential ${\cal V}$, which is consistent with the
selfenergy used and a calculation of the $\rho$-selfenergy in the nucleus
rather than in nuclear matter.


\section*{Acknowledgements}

This work was supported by BMBF, GSI Darmstadt and DFG.


\end{document}